\documentclass[conference]{IEEEtran}

\IEEEoverridecommandlockouts

\usepackage{latexsym}
\usepackage{amsfonts}
\usepackage{amsmath}
\usepackage{amssymb}
\usepackage{color}
\usepackage{epsfig}
\usepackage{xspace}
\usepackage{graphicx}
\usepackage{cite}
\usepackage{balance}
\usepackage{tablefootnote}
\usepackage{booktabs}
\usepackage{comment}
\usepackage{epstopdf}
\usepackage{multirow}
\usepackage{url}
\usepackage{array}
\usepackage{booktabs}
\usepackage{multirow}
\usepackage{caption}
\usepackage{subcaption}
\usepackage{latexsym}
\usepackage{amsfonts}
\usepackage{amsmath}
\usepackage{color}
\usepackage{epsfig}
\usepackage{xspace}
\usepackage{graphicx}
\usepackage{cleveref}
\usepackage{balance}
\usepackage{hhline}
\usepackage{float}
\usepackage{multirow}
\usepackage{listings}
\usepackage[table,xcdraw]{xcolor}
\usepackage{listings}
\usepackage{color}
\usepackage{caption}
\usepackage{subfig}
\usepackage{makecell}
\usepackage{marvosym}

\definecolor{dkgreen}{rgb}{0,0.6,0}
\definecolor{gray}{rgb}{0.5,0.5,0.5}
\definecolor{mauve}{rgb}{0.58,0,0.82}

\usepackage{algpseudocode}
\usepackage{newfloat}
\DeclareFloatingEnvironment[fileext=frm,placement={!ht},name=Frame]{myfloat}
\usepackage{listings}
\usepackage{epsfig}
\usepackage{multirow}
\usepackage{url}

\usepackage{enumitem}
\setlist{topsep=0pt,noitemsep} \setitemize[1]{label=$\circ$}
\newcommand{\myhrule}{\rule[.5pt]{\hsize}{.5pt}}

\newcommand{\eat}[1]{}

\newcommand{\bi}{\begin{itemize}}
\newcommand{\ei}{\end{itemize}}
        {\end{itemize}} %\vspace{-0.5ex}}

\newcommand{\mat}[2]{{\begin{tabbing}\hspace{#1}\=\+\kill #2\end{tabbing}}}

\newcommand{\be}{\begin{enumerate}}
\newcommand{\ee}{\end{enumerate}}
\newcommand{\beqn}{\begin{eqnarray}}
\newcommand{\eeqn}{\end{eqnarray}}

\newcommand{\stitle}[1]{\vspace{1.6ex}\noindent{\bf #1}}
\newcommand{\etitle}[1]{\vspace{1ex}\noindent{\underline{\em #1}}}

\newcommand{\ie}{\emph{i.e.,}\xspace}
\newcommand{\eg}{\emph{e.g.,}\xspace}

\newcommand{\refalg}[1]{Algorithm \ref{#1}}

\newcommand{\refeq}[1]{Eq \ref{#1}}
\newcommand{\reftab}[1]{Table \ref{#1}}

\newcommand{\Do}{\mbox{\bf do}\ }

\newcommand{\kw}[1]{{\ensuremath {\mathsf{#1}}}\xspace}

\newcounter{ccc}
\newcommand{\bcc}{\setcounter{ccc}{1}\theccc.}

\newcommand{\icc}{\addtocounter{ccc}{1}\theccc.}

\newcommand{\eop}{\hspace*{\fill}\mbox{$\Box$}}     % End of proof

\newcounter{example}%[section]
\renewcommand{\theexample}{\arabic{example}}
\newenvironment{example}{
        \vspace{1.5ex}
        \refstepcounter{example}
        {\noindent\bf Example \theexample:}}{
        \eop\vspace{1.5ex}}

\newcommand{\nthesection}{\arabic{section}}

\newcounter{theorem}
\renewcommand{\thetheorem}{\arabic{theorem}}
\newcounter{prop}
\renewcommand{\theprop}{\arabic{theorem}}
\newcounter{lemma}
\renewcommand{\thelemma}{\arabic{theorem}}
\newcounter{cor}
\renewcommand{\thecor}{\arabic{theorem}}
\newcounter{definition}[section]
\renewcommand{\thedefinition}{\nthesection.\arabic{definition}}

\newcounter{alg}[section]
\renewcommand{\thealg}{\nthesection.\arabic{alg}}

\newcounter{arule}
\renewcommand{\thearule}{\arabic{arule}}

\newcounter{claim}
\renewcommand{\theclaim}{\arabic{claim}}

\renewcommand{\texttt}[1]{{\small\textsf{#1}}}

\definecolor{gray}{rgb}{0.5,0.5,0.5}

\renewcommand{\O}{{\mathcal O}}

\newcommand{\revise}[1]{#1}

        \newcommand{\mei}{\end{myitemize}\vspace{0.6ex}}

\makeatletter
\newcommand\figcaption{\def\@captype{figure}\caption}
\newcommand\tabcaption{\def\@captype{table}\caption}
\makeatother

\newcommand{\name}{\kw{name}}

\usepackage[ruled,vlined]{algorithm2e}

\newcommand{\tabincell}[2]{\begin{tabular}
		{@{}#1@{}}#2\end{tabular}}

\newcommand{\nnsort}{\kw{NN}-\kw{sort}}

\usepackage{enumitem}

\setlist{topsep=0pt,noitemsep} \setitemize[1]{label=$\circ$}
\title{
\revise{Deep Learning Service for Efficient Data Distribution Aware Sorting}}

\eat{
\author{
\IEEEauthorblockN{TO ADD}
\IEEEauthorblockA{
\textit{Affiliation}\\
email@email}}

}

\eat{
\author{\IEEEauthorblockN{
Xiaoke Zhu\IEEEauthorrefmark{2}\ \ \ 
        Qi Zhang\IEEEauthorrefmark{4} 
	}
	\IEEEauthorblockA{
		Beihang University\IEEEauthorrefmark{2}\ \ \ 
        Meta Platforms\IEEEauthorrefmark{4}   
        \\
		zhuxk@buaa.edu.cn\ \ \ 
        qizhang@meta.com
	}
}
}

\author{\IEEEauthorblockN{Xiaoke Zhu\IEEEauthorrefmark{2}\thanks{\IEEEauthorrefmark{2}work done while author was with Yunnan University}}
\IEEEauthorblockA{\textit{Beihang University} \\
China \\
zhuxk@buaa.edu.cn}
\and
\IEEEauthorblockN{Qi Zhang}
\IEEEauthorblockA{\textit{Meta Platforms} \\
USA \\
qizhang@meta.com}
\and
\IEEEauthorblockN{Wei Zhou*\thanks{*corresponding author}}
\IEEEauthorblockA{\textit{Yunnan University} \\
China \\
zwei@ynu.edu.cn}
\and
\IEEEauthorblockN{Ling Liu}
\IEEEauthorblockA{\textit{Georgia Institute of Technology} \\
USA\\
ling.liu@cc.gatech.edu}
}
\eat{
\numberofauthors{1}
\author{
}
}%EAT
\date{}

\eat{
\usepackage{etoolbox}
\makeatletter
\patchcmd{\@makecaption}
  {\scshape}
  {}
  {}
  {}
\makeatletter
\patchcmd{\@makecaption}
  {\\}
  {.\ }
  {}
  {}
\makeatother
\def\tablename{Table}
}

\begin{document}
\maketitle % for IEEEtran

\begin{abstract}
In this paper, we present a neural network-enabled data distribution aware sorting method, coined as {\nnsort}. Our approach explores the potential of developing
deep learning techniques to speed up large-scale sort operations, enabling data distribution aware sorting as a
deep learning service. Compared to traditional pairwise comparison-based sorting algorithms, which sort data
elements by performing pairwise operations, {\nnsort} leverages the neural network model to learn the data
distribution and uses it to map large-scale data elements into ordered ones. Our experiments demonstrate the significant advantage of using {\nnsort}. Measurements
on both synthetic and real-world datasets show that {\nnsort} yields 2.18$\times$ to 10$\times$ performance improvement
over traditional sorting algorithms.
\end{abstract}

%\maketitle % for acmart

%%%%%%%% Section 1 %%%%%%%%
\section{Introduction}
%\vspace{-0.7ex}
\label{sec-intro}
Sorting is a fundamental operation in the realm of big data, powering everything from database systems~\cite{DBLP:journals/csur/Graefe06} to bigdata analysis~\cite{10.1093/bioinformatics/bts440}. As the scale of data continues to grow, traditional sorting algorithms face increasing limitations in performance. While methods such as Quick Sort family~\cite{GPU-quick1}, Merge Sort family~\cite{andersson1998sorting}, and Radix Sort family~\cite{GPU-radix,bit-RADIX-sort} have long been relied upon,  their comparison-based and non-comparison-based optimizations appear to be reaching its bottleneck.

Recent research~~\cite{zhu2021dlb,DBLP:journals/access/XiangZCCLZ19,TimKraska,DBLP:conf/cidr/KraskaABCKLMMN19} has extensively explored how deep learning models can enhance the performance of traditional big data systems and algorithms. 
For instance, Tim Kraska et al. introduced a learned index~\cite{TimKraska} that leverages an empirical cumulative distribution function (CDF) to improve the performance of traditional data structures. They also proposed a learned approach~\cite{DBLP:conf/cidr/KraskaABCKLMMN19,SageDB} to improve sorting performance.

Their method, known as \kw{SageDB} \kw{Sort}, utilizes a learned model to generate a roughly ordered state of elements by predicting (mapping) the positions of elements, and then refined it by a traditional sorting algorithm like Quick Sort. However, this approach has limitations. Conflicts often arise when converting the learned model's output, such as multiple elements being mapped to the same position, leading to performance bottlenecks (see Section \ref{sec-expt}). Resolving these conflicts~(especially when numerous) can be time-consuming, making it less efficient than traditional algorithms.

\eat{
Recently, 
many work~\cite{zhu2021dlb,DBLP:journals/access/XiangZCCLZ19,TimKraska,DBLP:conf/cidr/KraskaABCKLMMN19} has explored how deep learning models can enhance the performance of traditional bigdata systems and algorithms. 
For example, Tim Kraska et al. introduced a learned hash-model index that leverages an empirical cumulative distribution function (CDF) at a reasonable computational cost~\cite{TimKraska,DBLP:conf/cidr/KraskaABCKLMMN19}. While they also proposed using a CDF model to improve sorting performance~\cite{DBLP:conf/cidr/KraskaABCKLMMN19}.
In principle, their approach uses a learned model to sort unordered elements into a roughly ordered state, followed by a traditional sorting algorithm, like Quick Sort, to refine the results. However, this method has limitations. Significant conflicts often arise when converting the outputs of a CDF model, leading to performance issues. Resolving these conflicts using traditional sorting methods can be time-consuming, particularly when the conflicts are numerous and severe, ultimately making the performance worse than traditional algorithms.
}

The question of how to design an effective deep learning-based sorting algorithm remains unanswered. Specifically, key issues include determining which type of neural network performs best for sorting, understanding the complexity of neural network-based sorting, dealing with conflicts, and minimizing operations such as data comparison and movement during the sorting process.

To address these issues, this work presents \nnsort, a neural network-based sorting algorithm designed to move beyond traditional sorting paradigms. Instead of relying solely on comparisons or inherent data characteristics, \nnsort harnesses the power of neural networks to create a data distribution-aware sorting method. By training a model on historical data to predict the sorted positions of new data, \nnsort offers a novel approach that achieves efficient and scalable sorting while incorporating an effective conflict-handling mechanism.

\begin{figure*}[ht]
    \centering
	\begin{minipage}[t]{0.9\linewidth}
		\centering
		\includegraphics[width=1\linewidth]{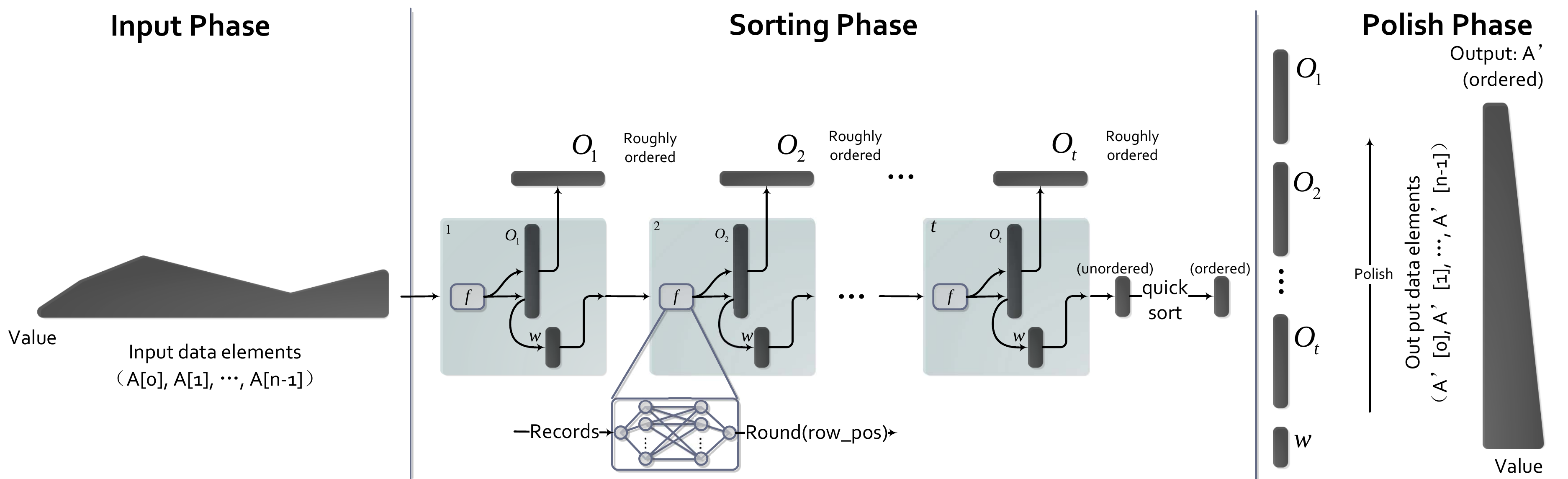}
		\caption{\nnsort architecture}
		\label{fig:framework}
	\end{minipage}
\end{figure*}

The architecture of \nnsort consists of three distinct phases: the input phase, where data is transformed into neural network-compatible vectors; the sorting phase, where a learned neural network iteratively refines the data order; and the polish phase, where traditional methods finalize the sorting (\ie ensuring the output is correct). This layered approach enables \nnsort to handle large datasets efficiently, minimizing the computational overhead from conflicts—a primary performance bottleneck in \kw{SageDB} \kw{Sort}. Moreover,
we systematically explored the potential of \nnsort, discussed its complexity, performance, and advantages over traditional algorithms. Through rigorous experiments on both synthetic and real-world datasets, we justify the effective of \nnsort.

The contributions of this paper are summarized as follows:
(a) we investigate the opportunities and challenges of enhancing traditional sorting processes by leveraging neural network-based learning approaches;
(b) we develop \nnsort, a novel neural network-based sorting method that utilizes historical data to train a data distribution-aware model. This trained model performs high-performance sorting on incoming data iteratively, with an additional touch-up process to ensure the correctness of the final result. In contrast to state-of-the-art learned sorting methods, \eg \kw{SageDB} \kw{Sort}, \nnsort scales effectively by reducing conflicts during CDF mapping;
(c) we provide a formal analysis of \nnsort's complexity using a cost model that clarifies the intrinsic relationship between model accuracy and sorting performance.
(d) we evaluate \nnsort's performance on both synthetic and real-world datasets. Experimental results demonstrate that \nnsort achieves up to an order of magnitude speed-up in sorting time compared to state-of-the-art sorting algorithms.

The rest of the paper is organized as follows: the design of \nnsort is presented in Section~\ref{sec:nn-sort}. The complexity of \nnsort is discussed in Section~\ref{sec-analysis}. The experimental results are presented and analyzed in Section~\ref{sec-expt}. Related work is reviewed in Section~\ref{sec-related work}, and the paper is concluded in Section~\ref{sec-conclude}.

%\input{sec-background}
%\vspace{-1.4ex}
%%%%%%%% Section 3 %%%%%%%%
\section{Neural Network Based Sort}\label{sec:nn-sort}
In this section, we discuss the design of \nnsort, including how to use a neural network model for effective sorting, as well as how such a neural network model can be trained.

\stitle{Challenges.}
Sorting involves mapping elements from an unsorted state to a sorted state. Rather than relying on traditional comparison-based methods (\eg Quick Sort), this mapping can be achieved through a data distribution-aware model that takes each element as input and returns its expected position within the sorted array. Given an ideal function 
$f$ and distinct elements $x,y\in A$ such that 
$x<y$, the function would ideally satisfy $f(x)<f(y)$. 
While such a "magic" function may not always hold perfectly, it can still accelerate sorting, with greater accuracy leading to more effective acceleration. Before discussing the methods, several challenges must be addressed to ensure both effectiveness and accuracy.
\eat{
Assuming that magic function existing, we have $x,y\in A$ and $x\neq y$ such that  $x < y \to f(x) < f(y)$.
Of course, that magic function not always hold, however, we can still use it to accelerating sorting.}
(1) first, for correctness, the model must precisely reflect the order among input data elements, producing results consistent with those of traditional sorting algorithms.  However, it is impossible to have that  function $f$, especially if we train the model just based on samples from the input data;
(2) second, for effectiveness, the model ideally should sort large volumes of data in a single pass. Achieving this requires a model to be both complex and accurate enough to capture the exact order of all elements, which can result in significant training cost and inference cost. Thus, a balance between model accuracy and sorting performance is crucial.
(3) third, conflicts arise when multiple input elements are mapped to the same output position, \ie  $f(a) = f(b)$ where $a, b\in A$ and $a \neq b$. Effectively managing these conflicts is crucial for both the correctness and efficiency of the learned sorting approach. \kw{SageDB} \kw{Sort} addresses these collisions using traditional sorting algorithms, such as Quick Sort, which incurs computational overhead when collisions is too large. That is to said, there is still room for improvement.

\eat{
Third, conflicts arise when multiple input elements are mapped to the same output position, \ie $f(a) = f(b)$ where $a, b\in A$ and $a \neq b$. Effectively managing these conflicts is crucial for both the correctness and efficiency of the learned sorting approach.  \kw{SageDB} \kw{Sort}~\cite{DBLP:conf/cidr/KraskaABCKLMMN19} addresses such collisions using traditional sorting algorithms, such as quick sort, which incurs additional computational overhead. However as shown in our experiments, there is still room to improve.
}

\begin{figure}[t]
	\centering
	\begin{minipage}{1\linewidth}
		\myhrule
		\vspace{-1ex}
		\mat{0ex} {
			\noindent {\bf Procedure} {\tt \nnsort}\\	
			\noindent {\sl Input:\/} \= $A$: the array of data elements \\$f$: the learned model \\ $m$: the relaxation factor \\ $\epsilon$ the predefined iteration limit \\
            $\tau$ the predefined threshold \\
			{\sl Output:\/} \= $A'$: the array of data elements after sorted\\
			\noindent
			\bcc \hspace{0ex} {\tt $w \gets$ translate($A$)}\\
			\icc \hspace{0ex} {\tt init $O\gets \emptyset$, $i \gets 0$ }\\
			\icc \hspace{0ex} {\mbox{\bf while}\ } {\tt $0<i<\epsilon$ and count($w$) > $\tau$ \Do}\\
			\icc \hspace{4ex} {\tt init $o_i\gets \emptyset$, $c\gets \emptyset$}\\
            \icc \hspace{4ex} {\tt $ w\_pos$ $\gets w.map(f)$}\\
			\icc \hspace{4ex} {\mbox{\bf for} } {\tt $idx$ in count($w$) \Do}\\
			\icc \hspace{8ex} {\tt $pos$ $\gets$ round($w\_pos[idx]*m$)}\\
            \icc \hspace{8ex} {\mbox{\bf if} } {\tt $o_i$[$pos$] is empty \Do}\\
            \icc \hspace{12ex} {\tt $o_i[pos] \gets w[{idx}]$ }\\
            \icc \hspace{8ex} {\mbox{\bf else}}\\
            \icc \hspace{12ex} {\tt $c$.append($w[{idx}]$)}\\
			\icc \hspace{4ex} {\tt $O$.append($o_i$)}\\
			\icc \hspace{4ex} {\tt $w\gets c$}\\
			\icc \hspace{4ex} {\tt ++i}\\
			\icc \hspace{0ex} {\tt QuickSort($w$)}\\
			\icc \hspace{0ex} {\tt $A'\gets$ polish($O,w$)}\\
			\icc \hspace{0ex} {\tt \mbox{\bf Return} $A'$}\\
		}
        \vspace{-1ex}
		\myhrule       
		\caption{Algorithm \nnsort}\label{algorithm}
        \vspace{-4ex}
	\end{minipage}		
\end{figure}

\stitle{NN-sort Design.}
In response to these challenges, we designed an iterative neural network-based sorting method. Unlike \kw{SageDB} \kw{Sort}, which employs a complex model to sort data in a single round, our approach utilizes a simpler model to sort over multiple rounds. Each round generates a roughly sorted array, with conflicts carried forward to the next iteration. This process continues until the conflicts in that iteration fall below a predefined threshold or a number of iterations is reached. The final small  conflict array is then sorted using a traditional method like Quick Sort and merged with the roughly sorted arrays. After a final traversal to ensure correctness, a fully sorted result is obtained.
The benefits are two folds: (1) using a simpler model reduces both inference and training costs; (2) the learned model can be applied repeatedly, avoiding direct sorting of conflicting elements with traditional methods.

Figure~\ref{fig:framework} illustrates this approach, where the input array $A$ is sorted into $A'$. The process is divided into three phases: \emph{Input}, \emph{Sorting}, and \emph{Polish}. Figure~\ref{algorithm} details the workflow of \nnsort, with Line 1 addressing the input phase, Lines 2-15 covering the sorting phase, and Line 16 corresponding to the polish phase.

\etitle{Input Phase.}
The input phase prepares the data for the neural network by encoding it appropriately, ensuring compatibility for processing. For example, string-type data is converted into ASCII values. This encoding step is crucial, as it standardizes the data format and enables the neural network to interpret and process a wide variety of input types, such as integers, floating-point numbers, or categorical data, in a structured and efficient manner. 
For simplicity, we denoted such operations as $w\gets \kw{translate}(A)$~(line 1, Figure~\ref{algorithm}).
%Additionally, if the input consists of numerical values, techniques such as normalization or one-hot encoding may be applied to optimize the neural network’s performance by improving the scalability of the data.

\begin{figure}[t]
	\centering
	\begin{minipage}{1\linewidth}
		\myhrule
		\vspace{-1ex}
		\mat{0ex} {
			\noindent {\bf Procedure} {\tt polish($O$)}\\	
			\noindent {\sl Input:\/} \= $O=\{o_1, o_2, ...\}$: array of roughly sorted arrays. \\ $w$: strictly sorted array.\\
			{\sl Output:\/} \= $A'$: the array of data elements after sorted\\
			\bcc \hspace{0ex} {\tt $A'\gets w$}\\				
			\icc \hspace{0ex} {\mbox{\bf for} } {\tt $o_i\in O$ \Do}\\
			\icc \hspace{4ex} {\tt $A'\gets$ InsertSort($A'$, $o_i$)}\\
			\icc \hspace{0ex} {\tt \mbox{\bf Return} $A'$}\\				
		}
        \vspace{-5ex}
		\myhrule
		\caption{Algorithm polish}\label{alg:polish}
	\end{minipage}
    \vspace{-1ex}
\end{figure}

\etitle{Sorting Phase.}
In the sorting phase, a learned model $f$ iteratively organizes unordered data into approximately sorted arrays. 
First, the learned model $f$ maps each element to its expected position~(line 5, Figure~\ref{algorithm}). 
Then, $o_i$ stores elements of $w$ based on their value in $w\_pos$, where $i$ denotes the iteration number. If a collision occurs in $o_i$—where multiple elements map to the same position—only the first element is stored in $o_i$, while subsequent elements are placed in a temporary conflict array 
$c$~(line 8-11, Figure~\ref{algorithm}). In the following iteration, elements in $c$ are reprocessed by learned model $f$~(line 13, Figure~\ref{alg:polish}). This process continues until either a predefined maximum number of iterations $\epsilon$ is reached or the size of $c$ drops below a threshold $\tau$, at which point it is sorted using a traditional algorithm.

It is worth mentioning that, 
each element in $w\_pos$ represents the expected position of corresponding elements of $w$ within the final sorted array. Instead of using $w\_pos[idx]$, which is the direct output of $f$, we use $round(w\_pos[idx] * m)$, a rounded value, to represent the position of $w[idx]$. The reasons are two folds: (1) the outputs of $f$ are decimals while the positions need to be integers. (2)  with relaxation factor $m$ the input data elements can be mapped into a larger space, thereby reducing the number of conflicts. 
In addition, all conflicting data elements are stored in a conflict array $c$ and used as input to $f$ for the next iteration.
If the model $f$ does not perform effectively, \ie the conflicting array may never shrink below $\tau$ or may decrease too slowly, potentially resulting in higher overhead than traditional sorting algorithms. To prevent this, a threshold 
$\epsilon$ limits the maximum number of iterations. As we will show in the experimental section, $\epsilon=2$ or $\epsilon=3$  are good enough for accelerating sorting. There is a clearly decreased edge effect on the number of iterations.

\eat{
Each iteration concludes at line 14, yielding a roughly sorted array $o_i$ and an updated conflict array $c$.  If the model $f$ does not perform effectively, \ie the conflicting array may never shrink below $\tau$ or may decrease too slowly, potentially resulting in higher overhead than traditional sorting algorithms. To prevent this, a threshold 
$\epsilon$ limits the maximum number of iterations.}

\eat{
All the conflicting data elements are stored in a conflicting array $c$ and used as the input of $f$ for the next iteration. Each iteration ends at line 14, after which a roughly sorted array $o_i$ and a conflicting array $c$ are generated. As shown in line 3, the iterations end when the size of the conflicting array becomes no larger than a threshold $\tau$. Also note that in case the model $f$ is not working well, the size of the conflicting array may never become smaller than $\tau$ or it decreases very slowly, which ends up with an even larger overhead than using traditional sorting algorithms. In order to prevent this from happening, another threshold $\epsilon$ is used to limit the maximum number of iterations. 
}

\etitle{Polish Phase.}
The polish phase refines the roughly sorted arrays $O=\{o_1, o_2, ...\}$ to ensure the correctness of the output. Figure~\ref{alg:polish} outlines this process, where the arrays in $O$ are polished and merged with the strictly ordered array $A'$. The algorithm iterates over each array in $O$, merging them with $A'$ one by one.  Elements in $o_i$  are either appended or inserted into the result, depending on their order relative to $A'$~(\ie Insert Sort).

\eat{
The roughly sorted arrays $O$ is polished and merged with the strictly ordered array $w$ to produce the final output $A'$, ensuring it is fully ordered. The algorithm iterates over the arrays in $O$, merging them with $A'$ one by one. Elements of $o_i$ is compared and appended  or insert to the result. Here insert operation and append operation depend on whether the element in the order of A'.
}

\stitle{Remark.}
Since $o_i$	 is only roughly ordered, out-of-order elements are inserted into their correct positions in
$A'$, ensuring \nnsort's reliability despite potential errors from the learned model.
Though insertion is costlier than appending, it is limited to out-of-order elements. As model accuracy improves, the polish phase incurs acceptable overhead. Section~\ref{sec-analysis} discusses \nnsort's complexity, with experimental results showing few out-of-order elements, yielding nearly linear performance.

\eat{

After all the iterations, the last conflicting array $w$ is sorted by traditional sorting algorithms before being merged with the roughly sorted arrays $\{o_1, o_2, ...., o_k, ..., o_t\}$ in the polish phase. The polish phase also refines the roughly sorted arrays to ensure the correctness of the final output.
}

\eat{
Alogirthm~\ref{alg:polish} illustrates more details about the polish phase.
Roughly ordered arrays $\{o_1, o_2, ...., o_k, ..., o_t\}$ are polished and merged with the strictly ordered array $w$ to create the final ordered output $A'$, {\tt which is guaranteed to be strictly ordered.} The algorithm goes over all the arrays $o_i$ in $O$, and merges them with $w$ one by one. 
Each element in $o_i$ and $w$ is iterated, compared, and appended to the $result$(Line 6 - line 8).
However, there are exceptions. Since $o_i$ is a roughly ordered array, an element in $o_i$, if out of order, needs to be inserted to the correct location in $A'$ instead of being appended~(Line 11). 
This step guarantees the absolute reliability of \nnsort, even if the learned model $f$ makes a mistake.
The cost of {\tt insert} is higher than {\tt append}. However, $insert$ is required only for the out-of-order elements in $o_i$. Therefore, the more accurate the model $f$ is, the less overhead the $polish$ will incur. We discuss the complexity of the \nnsort in Section \ref{sec-analysis}. Our experimental results show that the amount of out-of-order elements created by \nnsort can be negligible, thus the performance of \nnsort is near linear.
}

\begin{figure}[t]
\centering
	\begin{minipage}[t]{0.95\linewidth}
		\centering
		\includegraphics[width=1\linewidth]{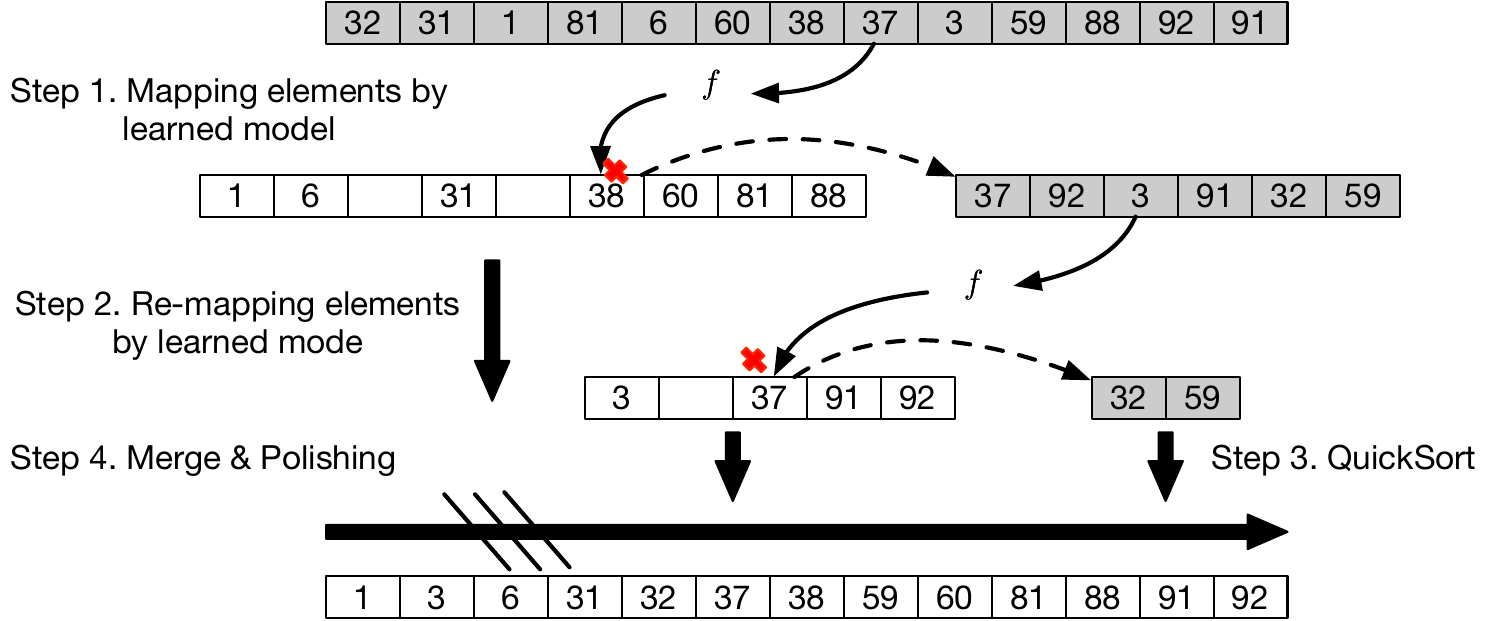}
		\caption{Example}
		\label{fig:example}
	\end{minipage}
\end{figure}

\begin{example}
Figure \ref{fig:example}  illustrates how \nnsort sorting.  

Given thresholds $\tau=2$, $\epsilon=2$ and an unordered array $A = \{32, 60, 31, 1, 81, 6, 88, 38, 3, 59, 37, 92, 91\}$, \nnsort first checks if $A$'s size is below $\tau$; if so, a traditional sorting method is applied. Otherwise, learned sorting begins.

Here, \nnsort processes $A$ in two rounds with a learned model. A conflict arises between elements 37 and 38, as $f$ maps them to the same position, placing 37 in a conflict array $c$. At the end of the first iteration, elements 92, 3, 91, 32, and 59 are also in the conflict array.

After the first iteration, since $w$'s size exceed $\tau$ and the iteration count is below $\epsilon$,  all elements in $c=\{37,92,3,91,32,59\}$ are reprocessed by $f$ in a second iteration, yielding a new sorted array $o_2=\{3, 37, 91,92\}$ and a smaller conflict array $c=\{32, 59\}$. 
Then a traditional sorting algorithm (\eg Quick Sort) is applied to $c$, and finally, $o_1, o_2$, and sorted $c$ are merged in the polish phase, producing a fully ordered result.

\eat{
Figure \ref{fig:example} 
provides a concrete example of how \nnsort orders a list of unsorted numbers. Consider $\tau=2$, Given a set of unordered data elements $A = \{32, 60, 31, 1, 81, 6, 88, 38, 3, 59, 37, 92, 91\}$, \nnsort first checks if the size of $A$ is smaller than a threshold $\tau$. If so, a traditional sorting method is applied; otherwise, the neural network-based sorting process begins.

In this example, $A$ is processed in two rounds with the help of the learned model. A conflict arises between elements 37 and 38, as $f$ maps them to the same position. Element 37 is stored in a conflict array $c$.

After the first iteration, since the size of $w$ (5) exceeds $\tau$ and the iteration count (1) is below the limit $\epsilon$, all elements in $w$ are fed into $f$ for a second iteration. This produces a new sorted array $o_2$ and a smaller conflict array $c$.   A traditional sorting method like Quick Sort is then applied to $w$. Finally, $o_1$, $o_2$, and the sorted $w$ are merged in the polish phase, yielding a fully ordered result.}
\end{example}

\stitle{Training.}
While training time is not the focus, all our models—whether shallow neural networks or simple linear/multivariate regression models—train quickly and perform well, as perfect position mapping (i.e., no conflicts or out-of-order elements) is unnecessary. The training and test data can differ; any learned order relationships help the model understand the sorting task.

\section{Model Analysis}
\label{sec-analysis}
This section establishes the time complexity of \nnsort by analyzing key operations—moving, mapping, and comparing data elements. A cost model is introduced to highlight relationships among factors like conflict rate, model scale, iteration count, out-of-order rate, data volume, and required operations.

The total operations of \nnsort is expressed as a  $T(n, e, \sigma, t, \theta)$, where: $n$ is the number of data elements to be sorted,
$e =  \{e_1, ..., e_t\}$ is the set of probabilities, with $e_i$ denoting the proportion of out-of-order elements in the $i$-th iteration,
$\sigma = \{\sigma_1,..., \sigma_t\}$ is the set of conflict rates, where $\sigma_i$ represents the conflict rate in the $i$-th iteration,
$t$ is the number of iterations completed,
$\theta$ denotes the number of operations required for each data element to pass through the neural network.
These basic notations are summarized in \reftab{notations}.

\begin{table}[t]
	\caption{Notations}
	\centering
	\begin{tabular}{ccc}
		\toprule  
		symbols&notations\\ 		
		\midrule
		$n$& the amount of data elements to be sorted\\
		%midrule 
		$\sigma_i$&  collision rate per iteration\\
		%\midrule 
		$e_i$& \tabincell{c}{ the number of data elements that were \\ out-of-order in the $i$-th iteration }\\
		%\midrule 
		$\epsilon$& the predefined limit of iterations\\
		%\midrule 
		$t$& the number of completed iterations\\
		%\midrule 
		$\theta$& \tabincell{c}{The operations required for data to pass through $f$ } \\
		%\midrule 		
		%$T(n, e, \sigma, t, \theta )$& \tabincell{c}{The total number of operations for \nnsort to sort $n$  elements.}\\
		\bottomrule 		
		%	$O={o_1, ..., o_t}$& \tabincell{c}{The intermediate  roughly ordered results generated by learned model $f$} \\
		%	\hline 	
	\end{tabular}
	\label{notations}
\end{table}

As shown in \refeq{eq:operations}, the number of operations for \nnsort to sort $n$ ($n>1$) data elements is $ C_1n^2+C_2nlogn+ C_3 n$.

\begin{equation}\label{eq:operations}
T(n, e, \sigma
		, t, \theta )=\left\{
\begin{array}{lr}
1, & if\ n=1 \\
C_1n^2+C_2nlogn+ C_3 n, & if\ n>1\\
\end{array}
\right.
\end{equation}

\begin{equation}
	\scriptsize
C_1 = [\frac{1}{2}\sum_{i=1}^{t}e_i(1-\sigma_i)(\prod_{j=1}^{i-1}\sigma_j)^2]  \notag
\end{equation}
\begin{equation}
	\scriptsize
C_2  = \prod_{j=1}^{t}\sigma_{j} \notag
\end{equation}

\begin{align}
	\scriptsize
C_3  = &  \sum_{i=1}^{t} [\theta \sum_{j=1}^{i}\sigma_{j} + (1-e_i)(1-\alpha_i)\prod_{j=1}^{i-1}\sigma_{j} +\prod_{j=1}^{i}\sigma_{j} ]\notag\\
 	 &  + (\prod_{j=1}^{t}\sigma_j)log(\prod_{j=1}^{t}\sigma_j) \notag
\end{align}

In \nnsort, the majority of the cost is spent in the Sorting and Polish phases. Let $s(n)$ represent the time spent in the Sorting phase and $p(n)$ represent the time spent in the Polish phase, we now formally analyze the complexity of \nnsort.

$s(n)$ consists of two kinds of operations: iteratively feeding the data elements into a learned model $f$ and sorting the array $w$ at the last iterations using traditional sorting algorithms~(\eg QuickSort), the time complexity of which is $nlogn$. If $n>1$, then $\theta\sum_{j=1}^{i}\sigma_{j}n$ operations are required to feed data into model $f$ in the $i$-th iteration. An additional $(\prod_{j=1}^{t}\sigma_j)nlog(\prod_{j=1}^{t}\sigma_j)n$ operations are required to keep $w$ order, since the size of  conflicting array $ w $ updated in the last iteration is $(\prod_{j=1}^{t}\sigma_j)n$. Therefore, at the end of the algorithm, the total operations of $s(n)$ is $(\prod_{j=1}^{t}\sigma_j)nlog(\prod_{j=1}^{t}\sigma_j)n+\theta\sum_{i=1}^{t}\sum_{j=1}^{i}\sigma_{j}n$.

\begin{equation}\label{eq:p1}
	\begin{aligned}
	T(n) & = s(n) +  p(n)  \text{\ , $(n>1)$}\notag\\
	= &(\prod_{j=1}^{t}\sigma_j)nlog(\prod_{j=1}^{t}\sigma_j)n+\theta\sum_{i=1}^{t}\sum_{j=1}^{i}\sigma_{j}n +p(n)\notag\\
	= &(\prod_{j=1}^{t}\sigma_j)nlog(\prod_{j=1}^{t}\sigma_j)n+\theta\sum_{i=1}^{t}\sum_{j=1}^{i}\sigma_{j}n\notag\\  
	& +\sum_{i=1}^{t}[e_i(1-\sigma_i)\prod_{j=1}^{i-1}\sigma_{j}n\times\frac{\prod_{j=1}^{i-1}\sigma_{j}n}{2} \notag\\
	& +(1-e_i)(1-\sigma_i) \prod_{j=1}^{i-1}\sigma_{j} n + \prod_{j=1}^{i}\sigma_{j} n]\notag\\	
	= &[\frac{1}{2}\sum_{i=1}^{t}e_i(1-\sigma_i)(\prod_{j=1}^{i-1}\sigma_j)^2] n^2 + \prod_{j=1}^{t}\sigma_{j} nlogn \notag\\
	  & + \{ \sum_{i=1}^{t} [\theta \sum_{n=1}^{i}\sigma_{j} + (1-e_i)(1-\alpha_i)\prod_{j=1}^{i-1}\sigma_{j} \\
        & +\prod_{j=1}^{i}\sigma_{j} ] 
	  + (\prod_{j=1}^{t}\sigma_j)log(\prod_{j=1}^{t}\sigma_j)\} n 
	\end{aligned}
\end{equation}

$p(n)$ involves two tasks: correcting any out-of-order elements and merging the intermediate arrays (\ie $o_1, ..., o_t$ and $w$). If no elements are out of order in $o_i$, \nnsort only needs to traverse the data once to merge them. However, in practice, out-of-order elements are almost inevitable, as the model $f$ is unlikely to be 100\% accurate. %As a result, both append and insert operations are required to merge these arrays and generate a fully sorted output.

For the ordered elements in $o_i$, \nnsort only requires appending it, with a time complexity of time complexity of $O(1)$.
Therefore, in the $i$-th iteration, at most $\prod_{j=1}^{i-1}\sigma_{j}n$ operations are required to complete the insertion, and at least one operation is needed to insert out-of-order elements. While, for an out-of-order element in the $i$-th merge iteration, $\frac{\prod_{j=1}^{i-1}\sigma_{j}n}{2}$ operations are required to insert it into the final ordered result.
Theoretically, assume that there are  $e_i(1-\sigma_i)\prod_{j=1}^{i-1}\sigma_{j}n$ out-of-order elements in the $i$-th iteration. It takes a total of $e_i(1-\sigma_i)\prod_{j=1}^{i-1}\sigma_{j}n \times \frac{\prod_{j=1}^{i-1}\sigma_{j}n}{2}$ operations to process these elements. Correspondingly, $(1-e_i)(1-\sigma_i)\prod_{j=1}^{i-1}\sigma_{j}n$ elements in $o_i$ and $\prod_{j=1}^{i}\sigma_{j}n$ elements in $w$ remain ordered. 
Thus in the $i$-th merge iteration, a total of $\prod_{j=1}^{i}\sigma_{j} n + (1-e_i)(1-\sigma_i) \prod_{j=1}^{i-1}\sigma_{j} n $ operations are required to append the ordered elements to the final result. Overall,  \nnsort requires a total of $\sum_{i=1}^{t}[e_i(1-\sigma_i)\prod_{j=1}^{i-1}\sigma_{j}n\times\frac{\prod_{j=1}^{i-1}\sigma_{j}n}{2}+(1-e_i)(1-\sigma_i) \prod_{j=1}^{i-1}\sigma_{j} n  + \prod_{j=1}^{i}\sigma_{j} n]$ to sort $n$ data elements~(We show the detail in Equation~\ref{eq:operations}). %We point that formal logic in \refeq{eq:p1}.

%%%%%%%% Section 7 %%%%%%%%
\section{Experimental Study}
\label{sec-expt}
Using real and synthetic data, we conducted five experiments to evaluate (1) overall sorting performance, (2) iteration impact, and (3) effects of changing data distribution.

%We also show (4)%cost breakdown analysis. 
%(5) evaluation of training steps to converge.

\begin{figure*}[t]
\centering
	\begin{minipage}[t]{1\linewidth}
		\centering
		\includegraphics[width=1\linewidth]{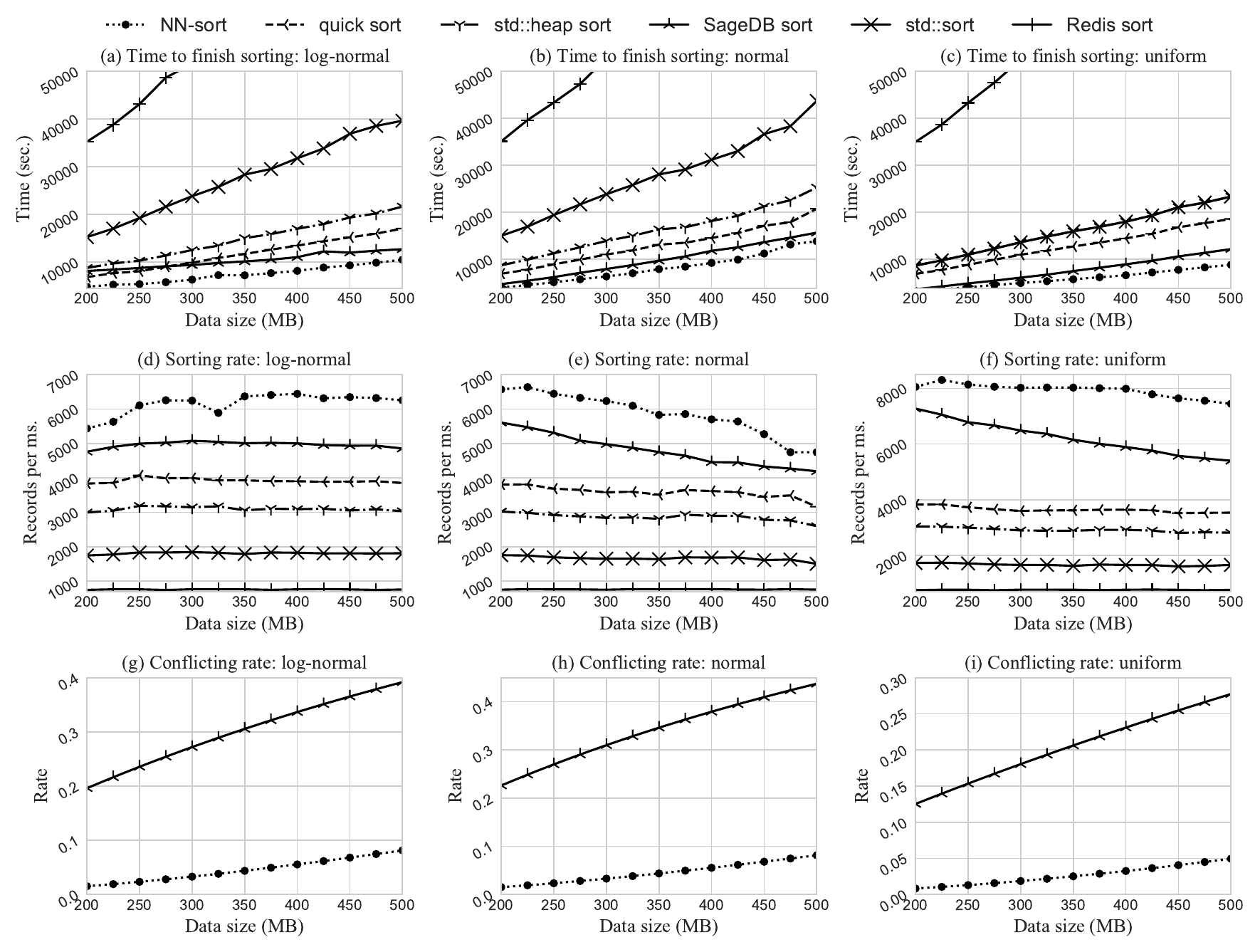}
		\caption{Overall performance evaluation}
		\label{fig:performance}
	\end{minipage} 
\end{figure*}

\subsection{Experimental setup}
\stitle{Datasets.}
 The datasets used in this section are generated from the most commonly observed distributions in the real world, such as uniform distribution, normal distribution, and log-normal distribution. The models used in the experiments are trained over a subset of the testing data. The sizes of the testing dataset vary from 200MB to 500MB and each data element is 64 bits wide floating number.
To verify the performance of the \nnsort under the real-world dataset. We use the QuickDraw game dataset from Google Creative Lab\cite{quick-draw}, which consists of $50,426,265$ records
 of schema \{'key-id', 'word', 'country code', 'timestamp', 'recognized', 'drawing'\}.  Sorting is perform on 'key-id'.
% and each records has 6 properties: \{'key-id', 'word', 'country code', 'timestamp', 'recognized', 'drawing'\}. 
%If no special declaration is made, all the predict model are trained over a subset of data elements which are independently and identically distributed with the test data. 

\stitle{Baselines.}
We  compared with five baselines
(1) \textbf{Quick Sort}~\cite{introcution-to-algorithm}: This algorithm divides the input dataset into two independent partitions, such that all the data elements in the first partition are smaller than those in the second partition. Then, the dataset in each partition is sorted recursively. The time complexity of \textbf{Quick Sort} can achieve $\O(nlogn)$ in the best case while $\O(n^2)$ in the worst case.    
(2) \textbf{std::sort}~\cite{C++}: std::sort is one of the most widely used sorting algorithms from c++ standard library, and its time complexity is $\O(nlogn)$
(3) \textbf{std::heap sort}~\cite{C++}: \textbf{std::heap sort} is another sorting algorithm from c++ standard library, and it guarantees to perform at $\O(nlogn)$ time complexity.
(4) \textbf{Redis Sort}~\cite{Redis}: Redis Sort is a sorting method based on a data structure named $sortSet$. To sort $M$ data elements in a $sortSet$ of size $N$, the efficiency of \textbf{Redis Sort} is $\O(N+Mlog(M))$.
In addition, we also compared \nnsort with (5) \textbf{SageDB Sort}~\cite{DBLP:conf/cidr/KraskaABCKLMMN19,SageDB}, leading performance DNN-based sorting method.%, the core idea of which is to speed up sorting by using a CDF model to organize the data elements in roughly sorted order, and then use traditional sorting algorithms to sort the data elements that are out of order. %Unlike our work, \textbf{SageDB Sort} maps data elements only once by CDF, which results in higher conflicting rate thus lower sorting efficiency.
The relaxation factor $m$ is set to 1.25 for learned sorting methods to reduce conflicts.

\stitle{Measurements.}
We used sorting time and sorting rate of Equation~\ref{eq:sorting rate} to evaluate the overall performance.
\begin{equation}\label{eq:sorting rate}
	\text{sorting rate} = \frac{\text{elements}}{\text{time to finish sorting}}
\end{equation}
We also used traditional sorting rate to evaluate learned-based sorting methods which is described in Equation~\ref{eq:ratio}.
This rate indicates how many data elements still require traditional sorting due to model inaccuracy in the learning-based approach. Ideally, a lower traditional sorting rate signifies the better performance of learning-based sorting.

\begin{equation}\label{eq:ratio}
    \text{Traditional sorting rate} = \frac{\text{size(last conflicting array $w$)}}{\text{size of the original array A}}
\end{equation}
 
\stitle{Environment.}
Experiments were conducted on a machine with 64GB RAM, a 2.6GHz Intel i7 processor, and a GTX1080Ti GPU with 16GB memory, running RedHat Enterprise Server 6.3. Each result reported is the median of ten runs.

\stitle{Training details.}
We employed a regression model with three hidden layers, containing 2, 6, and 1 neurons, respectively. A rounding function is used to determine each element’s final position. Adam~\cite{adam} was the chosen optimizer. The training was conducted using a GPU and is performed offline, so training time is excluded from the runtime.

\begin{equation}\label{eq:loss}
loss_\delta=\left\{
\begin{array}{lr}
\frac{1}{2}(f(x_i)-label_i)^2, & if\ |f(x_i)-label_i|\le \delta, \\
\delta|f(x_i)-label_i|-\frac{1}{2}\delta^2, & otherwise\\
\end{array}\right.
\end{equation}

To avoid the impact of outliers during training, the model used in experiments is trained according to the Huber loss \cite{Huber1964Robust} as shown in Equation~\ref{eq:loss}.
The batch size for training is set to 128. For all environments, we use the Adam optimizer with a learning rate of 0.001.

\begin{table}[t]	
	%\small
	\caption{Evaluation under real-world data}
	\label{tab:real-world-data}
	\centering
	\begin{tabular}{cccc}
		\toprule  
		\textbf{\tabincell{c}{Algorithm \\name}} & \tabincell{c}{ Time (sec.)} & \tabincell{c}{Sorting Rate \\(elements/sec.)}  & \tabincell{c}{The traditional \\ sorting rate\\(\%)}\\
        \midrule
		\tabincell{c}{Quick Sort} & $10.86$ & $4666.14$&-\\
		\tabincell{c}{std::heap sort} & $13.46$  & $3746.44$&-\\
		\tabincell{c}{std::sort} & $23.71$  & $2127.19$&-\\
		\tabincell{c}{Redis Sort} & $63.14 $  &  $798.6320 $ & -\\
		\tabincell{c}{SageDB Sort}& $10.53$  & $4790.125$&$9.16$\\
		\tabincell{c}{\nnsort}& $8.47$  & $5950.186$& $0.4$\\
		\bottomrule 
	\end{tabular}
\end{table}

\subsection{Experimental results.}

\stitle{Exp-1: Overall Sorting Performance.}\label{sec:sorting-performance}
Figure~\ref{fig:performance} presents a performance comparison of \nnsort against traditional sorting algorithms across various datasets with increasing sizes. Figures~\ref{fig:performance}(a)–(c) show the total sorting time, while Figures~\ref{fig:performance}(d)–(f) illustrate the sorting rates. Figures~\ref{fig:performance}(g)–(i) highlight the traditional sorting rate comparison between \nnsort and SageDB sort, as defined in Equation~\ref{eq:ratio}.
we observe the following:

\nnsort exhibits notable advantages over traditional sorting algorithms. As shown in Figure~\ref{fig:performance}~(d), its sort rate for a lognormal distribution dataset reaches nearly 8,300 elements per second, outperforming std::heap sort by 2.8$\times$, Redis Sort by 10.9$\times$, std::sort by 4.78$\times$, and Quick Sort by 218\%. It also exceeds SageDB Sort by 15\%.
The dataset’s value range—defined by its maximum and minimum values—affects \nnsort’s performance. As shown in Figure~\ref{fig:performance}~(h), a slight decline in sorting rate occurs with highly concentrated values, which create more conflicts and reduce efficiency. In contrast, fewer records within the same range enhance sorting performance.

\nnsort achieves optimal performance with uniformly distributed data, reaching a sorting rate of approximately 8,000 records per second—about 1.3$\times$ higher than with a normal distribution—due to fewer conflicts in uniformly distributed records.

Compared to SageDB Sort, \nnsort consistently reduces reliance on traditional sorting. A larger proportion of elements are accurately sorted by \nnsort's neural model, minimizing the need for the more time-consuming traditional sorting and contributing to \nnsort’s superior performance over SageDB Sort.

\eat{
It can be clearly seen that.
(1) \nnsort has significant performance advantages over traditional sorting algorithms. For example, Figure~\ref{fig:performance}(d) shows that the sort rate of \nnsort for the dataset with lognormal distribution is almost 8300 data elements per second, which is 2.80 times of std::heap sort, 10.90 times of Redis Sort, 4.78 times of std::sort, 218\% higher than Quick Sort, and also outperforms SageDB Sort by 15\%. 
(2) The range, \ie the maximum and minimum value of the dataset affects the performance of \nnsort, e.g. there has been a slight decline in sorting rate in Figure~\ref{fig:performance}(h). This is because the concentrated value of records leads to more conflicts and these conflicts will degrade the performance. In other words, the fewer records within the same range, the better the sorting performance.
(3) NN sorting performs better with uniform data distribution than with other distributions. The sorting rate is about 8000 records per sec with uniform data distribution, which is about $1.3\ times$ with normal data distribution. The reason is that records generated from a uniform distribution have fewer conflicts.
(4) \nnsort is able to consistently maintain a much lower traditional sorting rate compared with SageDB Sort. In other words, more elements can be correctly sorted by the model in \nnsort than that in SageDB sort, thus less efforts need to be made by the traditional sorting which is more time-consuming. This also explains why \nnsort performs better than SageDB sort.
}

\stitle{Exp-2: Evaluation on real-word Dataset.}\label{exp:real-word}
Using the model trained in previous sections on uniformly distributed data, we evaluated \nnsort’s performance on the real-world dataset QuickDraw. As shown in Table~ \ref{tab:real-world-data}, \nnsort delivers significant performance gains over traditional sorting algorithms on real-world data. With a sorting rate of 5,950 elements per second, \nnsort outperforms std::sort by 2.72$\times$ and Redis Sort by 7.34$\times$, and is also 58\% faster than std::heap sort. Additionally, \nnsort surpasses SageDB Sort in both traditional sorting rate and overall sorting rate.

\eat{
We used the model that is trained in previous subsections under uniformly distributed data to evaluate the performance of \nnsort over real-world dataset, QuickDraw.
As shown in Table \ref{tab:real-world-data}, \nnsort shows significant performance benefits over traditional sorting under real-world data. In terms of the sorting rate, \nnsort is 5950 per second, which is 2.72 times of std::sort and 7.34 times of Redis Sort. It is also 58\% faster than std::heap sort. We can also observe that \nnsort outperforms SageDB Sort in terms of both the traditional sorting rate and sorting rate.
}

%, and the data distribution(including both training data and sorting data) will affect the time to finish sorting. As shown in Figure~\ref{fig:performance-step:normal}, \nnsort spends longer time on sorting dataset with normal distribution, since more conflicts are created in this scenario. Therefore, the fewer conflicts per iteration, the better \nnsort can perform.

\begin{figure}[t]
	\centering
	\begin{minipage}[t]{0.9\linewidth}
		\centering
		\includegraphics[width=1\linewidth]{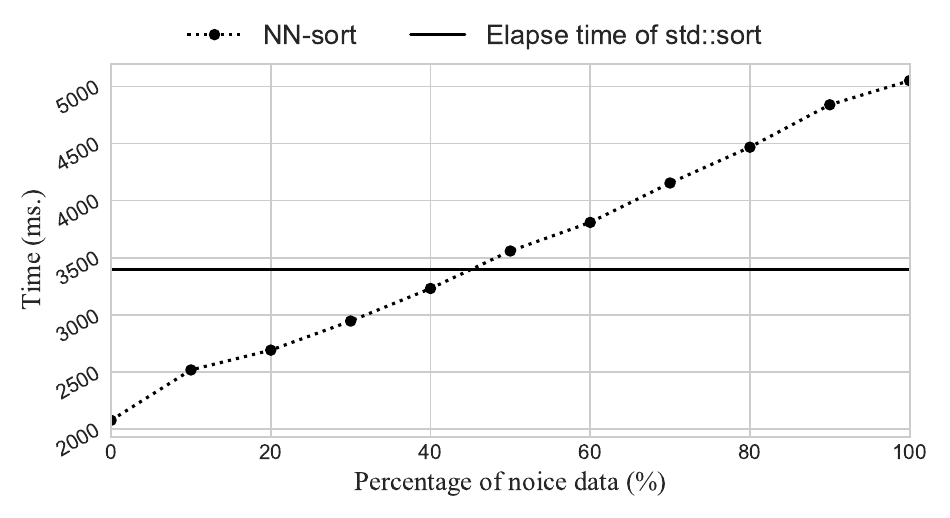}
		\caption{The impact of data distribution on \nnsort performance}
		\label{fig:notice}
	\end{minipage} 
\end{figure}

\begin{figure*}[t]
	\centering
	\begin{minipage}[t]{1\linewidth}
		\centering
		\includegraphics[width=1\linewidth]{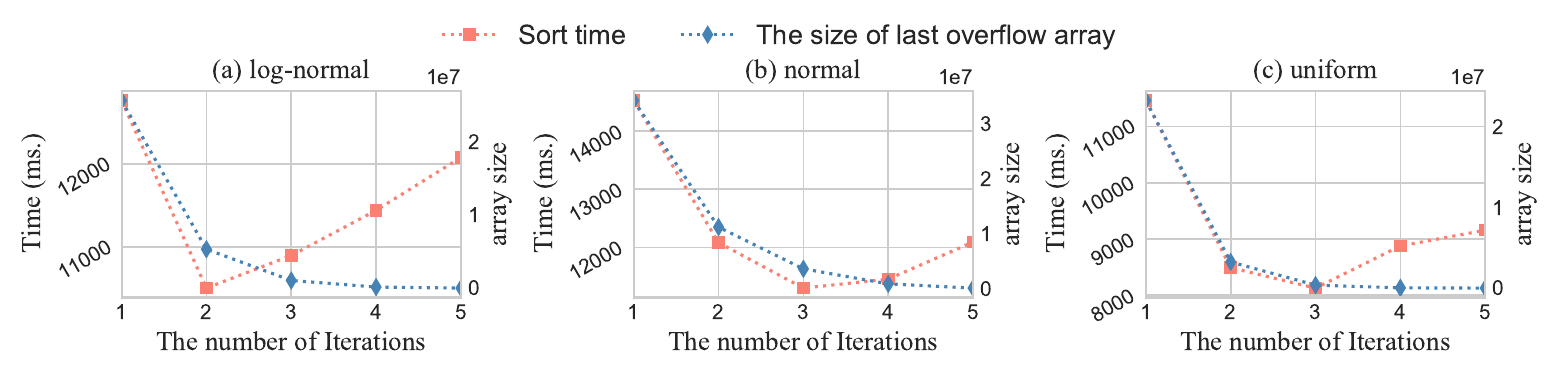}
		\caption{Impact of Iterations}
		\label{fig:Iterations}
	\end{minipage} 
\end{figure*}

\stitle{Exp-3: Impact of the Changing Data Distribution.}\label{exp:impact of distribution}
As shown in previous experiments, \nnsort performs optimally when the distribution of the sorting data resembles that of the training data. But how does it perform when faced with a different data distribution? To explore this, we trained a model using a 100MB uniformly distributed dataset, then applied it to sort datasets with varying distributions. Specifically, the test dataset combined uniformly and normally distributed data, with the latter considered "noisy." Sorting time was measured to assess \nnsort’s effectiveness, as shown in Figure~\ref{fig:notice}. Results indicate that as noise in the dataset increases, \nnsort's effectiveness declines due to a growing number of conflicts elements generated in each iteration when the test data distribution diverges from the training data. These elements must then be handled by traditional algorithms during the polish phase. Nonetheless, \nnsort shows resilience to distributional changes, outperforming the widely-used std::sort algorithm even with up to 45% noisy data.

\eat{
As discussed in the previous experiments, \nnsort works well when the sorting data is in the same or similar distribution as the training data. A natural question to ask is what if the sorting data has a different distribution than the training data? To answer this question, we trained a model by using a dataset that contains 100MB of uniformly distributed data elements. Then, we use this model to sort datasets with different distributions. Specifically, the dataset to be sorted is a mix of data with uniform and normal distributions, and we denote the normal distribution data as noisy data. The sorting time is measured to reflect the effectiveness of \nnsort and the results are displayed in Figure \ref{fig:notice}. It can be observed that the effectiveness of \nnsort decreases as the dataset becomes more noisy. This is because when the distribution similarity between the training data and sort data decreases, more out-of-order data elements are produced by \nnsort in each sorting iteration, which need to be sorted by traditional sorting algorithms in the polish phase. However, \nnsort can tolerate data distribution change to some extent. As shown in Figure \ref{fig:notice}, \nnsort can still outperform one of the fastest and widely used sorting algorithms - std::sort - with up to 45\% noisy data.
}

\stitle{Exp-4: Impact of Iterations.}
\nnsort’s sorting performance is influenced by both the size of the final conflicting array and the number of iterations. Increasing the number of iterations reduces the size of the remaining conflicting array that requires traditional sorting, yet also extends model processing time. Conversely, fewer iterations leave a larger conflicting array, increasing the time required for traditional sorting. In this set of experiments, we quantify these factors to guide users in optimizing \nnsort for improved performance.

In Figure~\ref{fig:Iterations}, the rhombus-dotted line represents the size of the final conflicting array, while the round-dotted line indicates total sorting time. Results show that while additional iterations reduce the size of the final conflicting array, they don’t necessarily improve performance, as each iteration requires the model to process all input data elements. Our experiments suggest that 2-3 iterations provide an optimal balance between conflicting array size and sorting time.

\eat{
The sorting performance can be affected by the size of the last conflicting array as well as the number of iterations. Concretely speaking, if the number of iterations increases, the elements of the last conflicting array which needs to be sorted using traditional methods, will decrease, but the time spent on the model will become longer due to more iterations. On the contrary, if the number of iterations is small, the size of the last conflicting array can be large, which requires a long time to be sorted by the traditional sorting algorithm. In this set of experiments, we quantify how these two factors can affect the performance of \nnsort, so as to guide practitioners or researchers to make a more informed decision on how to tune \nnsort for better performance. 

In Figure~\ref{fig:Iterations}, the rhomb-dotted line represents the size of the last conflicting array, while the round-dotted line illustrates the sorting time. It shows that the more iterations \nnsort has, the smaller the size of the last conflicting array is. However, this does not mean that the more iterations, the better performance of \nnsort. This is because each iteration needs to invoke the model multiple times, which equals to the number of input data elements. It can be observed from the experiments that $2-3$ iterations can be good enough. 
}

\begin{figure}[t]
	\centering
	\begin{minipage}[t]{1\linewidth}
		\centering
		\includegraphics[width=0.9\linewidth]{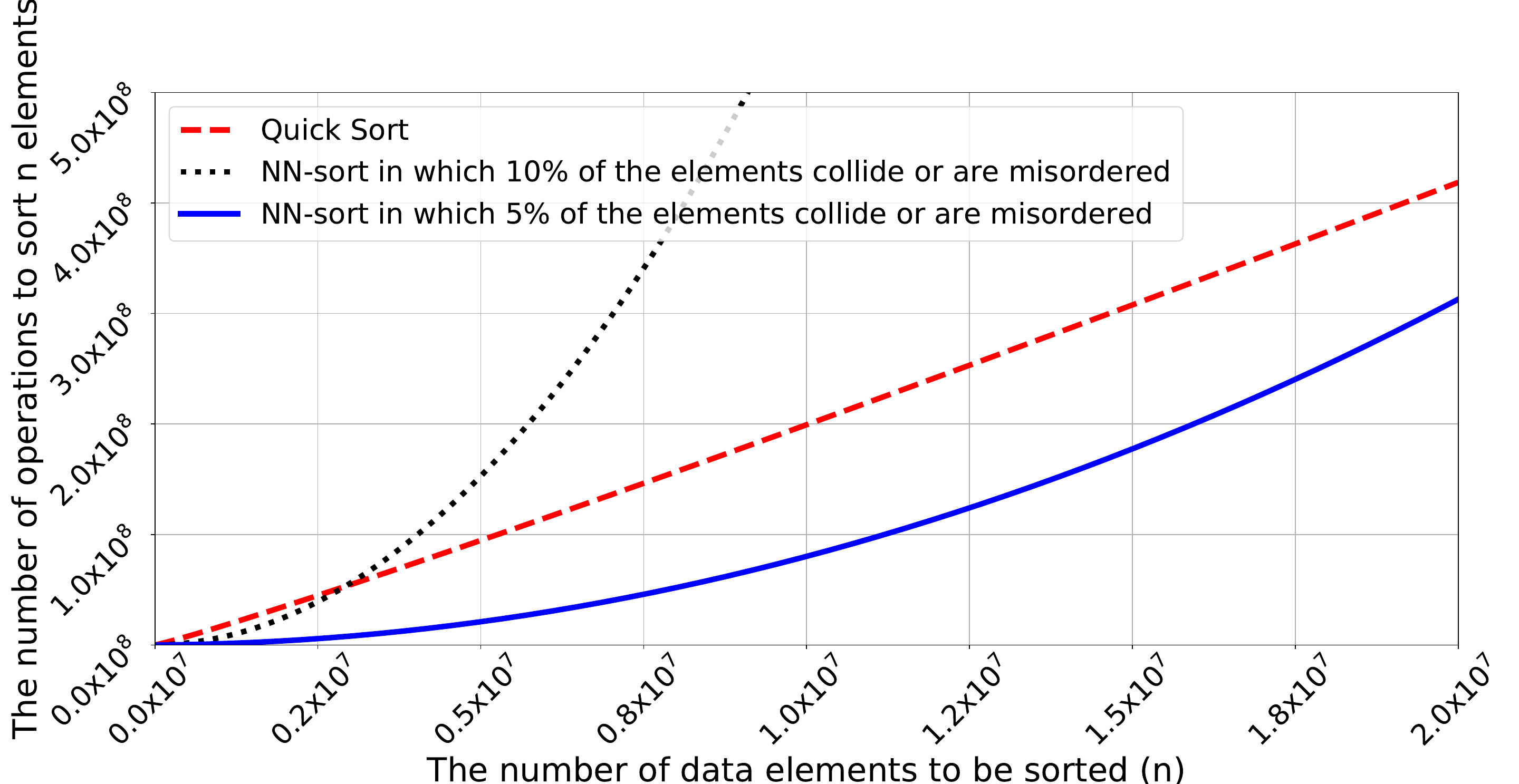}
		\caption{Comparison of operations between traditional sorting algorithm and \nnsort with different model qualities.}
		\label{fig:cost model}
	\end{minipage} 
\end{figure}

\stitle{Exp-5: Evaluation of sorting accuracy.}
A more complex neural network generally enhances model expressibility, resulting in lower conflict rates, and fewer out-of-order elements, but higher inference times. Thus, understanding the impact of these factors on \nnsort's overall time complexity is essential. In this section, we use a cost curve to illustrate how model quality—represented by the conflict rate $\sigma_i$ and out-of-order rate $e_i$ in each sorting iteration—affects \nnsort’s performance.

Figure~\ref{fig:cost model} compares the number of operations required by \nnsort to sort $n$ elements against Quick Sort’s baseline complexity ($\O(n\log n)$). To illustrate performance variations, we adjust \nnsort’s model quality. This analysis assumes \nnsort performs up to five iterations ($t=5$), with a model scale $\theta$ of 32 neurons, and equal conflict and out-of-order rates ($\sigma_i = e_i$) that remain constant across iterations. Results show that \nnsort substantially outperforms Quick Sort when conflict and out-of-order rates are at 10\%, with even greater performance gains as these rates drop to 5%.

\eat{
Figure~\ref{fig:cost model} shows the number of operations required by \nnsort to sort $n$ elements compared to Quick Sort ($\O(n\log n)$) as a baseline. We vary the model quality in \nnsort to demonstrate performance changes. The analysis assumes that \nnsort executes a maximum of five iterations ($t=5$), the model has a scale $\theta$ of 32 neurons, and the conflict rate ($\sigma_i$) equals the out-of-order rate ($e_i$), remaining constant across iterations. The results indicate that \nnsort significantly outperforms Quick Sort when the conflict and out-of-order rates are 10\%, and performance improves further when these rates drop to 5\%.
}

In summary, fewer conflicts and misordered elements result in more efficient sorting with \nnsort. A well-trained model with a misorder rate of 10\% or lower can outperform traditional sorting algorithms in terms of computational efficiency.

\section{Related Work}\label{sec-related work}
Sorting is one of the most fundamental algorithms in computing. We identify two key research areas: methods to reduce sorting time complexity and neural network-based data structures.

\stitle{Methods for reducing the sorting time complexity.} 
Many researchers have focused on accelerating sorting by reducing its time complexity. 
Traditional comparison-based sorting algorithms like 
Quick Sort, Merge Sort, and Heap Sort require at least 
$logn!\approx nlogn-1.44n$ 
operations to sort n elements~\cite{DBLP:conf/alenex/EdelkampW19}. 
Among these, Quick Sort achieves 
$\O(nlogn)$ average complexity but degrades to $\O(n^2)$ 
in the worst case. Merge Sort, while guaranteeing a worst-case of 
$nlogn-0.91n$, 
requires additional linear space relative to the number of elements \cite{DBLP:conf/alenex/EdelkampW19}. 
To mitigate the drawbacks of these algorithms and further reduce sorting time, researchers have combined different sorting techniques to leverage their strengths.
Tim Sort \cite{Python}, the default sorting algorithm in Java and Python, combines Merge Sort and Insertion Sort \cite{introcution-to-algorithm} to achieve fewer than  $nlogn$ comparisons on partially sorted arrays.
Stefan Edelkamp et al. proposed Quickx Sort \cite{DBLP:conf/csr/EdelkampW14}, which uses at most 
$nlogn-0.8358n+\O(logn)$ operations for in-place sorting. They also introduced a median-of-medians variant of Quick Merge Sort~\cite{DBLP:conf/alenex/EdelkampW19}, which employs the median-of-medians algorithm for pivot selection, reducing the operation count to  $nlogn+1.59n+\O(n^{0.8})$.

Redis Sort \cite{Redis} is a build-in sorting method of the Redis database based on the sortSet data structure. It sorts M elements in a sortSet of size 
N with an efficiency of $\O(N+Mlog(M))$.

Unlike previous work, this approach uses a learned model complexity to map an unordered array to a roughly ordered state, reducing overall operations. In the worst case, \nnsort has  complexity $\O(n^2)$ if all elements map to the same position, though practical operations remain lower than traditional sorting, 
This is validated by our experiment in Figure~\ref{fig:cost model}.

\eat{
Different from previous work, this work leverages a learned model, with a complexity of $\O(n)$ for mapping an unordered array to a roughly ordered one, to reduce the overall operations. In the worst case, \nnsort has a complexity of $\O(n^2)$—when all elements are mapped to the same position by the model. However, in practice, the number of operations is significantly lower than traditional sorting algorithms, provided the model is sufficiently accurate. This is validated by our experiment in Figure~\ref{fig:cost model}.
}

\stitle{Learned data structures and algorithms.}
This thread of research is to explore the potential of using the neural network-based learned data structures to improve the performance of systems. Tim Kraska \cite{DBLP:conf/cidr/KraskaABCKLMMN19,TimKraska} discussed the benefits of learned data structures and suggested that R-tree can be optimized by learned data structures. Xiang et al. \cite{DBLP:journals/access/XiangZCCLZ19} proposed an LSTM-based inverted index structure. By learning the empirical distribution function, their learned inverted index structure led to fewer average look-ups when compared with traditional inverted index structures. Alex Galakatos et al. \cite{DBLP:conf/sigmod/GalakatosMBFK19} presented a data-aware index structure called FITing-Tree, which can approximate an index using piece-wise linear functions with a bounded error specified at construction time. Michael Mitzenmacher \cite{DBLP:journals/corr/abs-1901-00902} proposed a learned sandwiching bloom filter structure, while the learned model is sensitive to data distributions. 

Unlike the research mentioned above, our approach integrates sorting with learning by training a model to enhance sorting performance. Additionally, we employ an iteration-based mechanism to further optimize performance by minimizing conflicts. We also provide a formal analysis of the time complexity of our approach and present a cost model to balance model accuracy with sorting performance.
A closely related work is SageDB Sort\cite{DBLP:conf/sigmod/GalakatosMBFK19,SageDB}, which leverages deep neural networks for sorting. Our approach improves upon SageDB Sort by offering a more efficient solution for handling position conflicts generated by the learned model.
%%%%%%%% Section 8 %%%%%%%%
\section{Conclusion}
\label{sec-conclude}

Sorting is fundamental in big data processing. We introduce \nnsort, a neural network-based sorting method that uses historical data to sort new data, iteratively reducing sorting conflicts—a key bottleneck in learned sorting. Our analysis includes complexity, a cost model, and the balance between model accuracy and performance. Experiments show \nnsort outperforms traditional algorithms. Future work includes enhancing \nnsort's adaptability to changing data distributions.

\eat{
\section*{Acknowledgement}
We sincerely thank Jing He, Shaowen Yao, and  Taining Cheng for their comments and helpful discussions. 

This work was supported by the National Natural Science Foundation of China under Grant 62162067 and 62101480, in part by the Yunnan Province expert workstations under Grant202305AF150078.
}
%This work was supported by the National Key R\&D Program of China (2021ZD0113903),  National Natural Science Foundation of China (No. U23B2056),  NSFC 62202313,Guangdong Basic and Applied Basic Research Foundation 2022A1515010120.

%%%%%%%%%%%%%%%

%\section*{Acknowledgment}
%This work is supported in part by The National Key Research and Development Program of China (2016YFB1000103) and NSFC (61602023, 61421003).

\balance
%\vspace{1ex}

%\eat{%EAT
%\renewcommand{\baselinestretch}{0.99}
%\bibliographystyle{abbrv}
\bibliographystyle{IEEEtran}
%\bibliographystyle{unsrt} % for IEEEtran
%\bibliographystyle{ACM-Reference-Format}
%\vspace{-0.5ex}
%\begin{small}
\bibliography{paper}

% Generated by IEEEtran.bst, version: 1.12 (2007/01/11)
\begin{thebibliography}{10}
\providecommand{\url}[1]{#1}
\csname url@samestyle\endcsname
\providecommand{\newblock}{\relax}
\providecommand{\bibinfo}[2]{#2}
\providecommand{\BIBentrySTDinterwordspacing}{\spaceskip=0pt\relax}
\providecommand{\BIBentryALTinterwordstretchfactor}{4}
\providecommand{\BIBentryALTinterwordspacing}{\spaceskip=\fontdimen2\font plus
\BIBentryALTinterwordstretchfactor\fontdimen3\font minus \fontdimen4\font\relax}
\providecommand{\BIBforeignlanguage}[2]{{%
\expandafter\ifx\csname l@#1\endcsname\relax
\typeout{** WARNING: IEEEtran.bst: No hyphenation pattern has been}%
\typeout{** loaded for the language `#1'. Using the pattern for}%
\typeout{** the default language instead.}%
\else
\language=\csname l@#1\endcsname
\fi
#2}}
\providecommand{\BIBdecl}{\relax}
\BIBdecl

\bibitem{DBLP:journals/csur/Graefe06}
G.~Graefe, ``Implementing sorting in database systems,'' \emph{{ACM} Comput. Surv.}, vol.~38, no.~3, p.~10, 2006.

\bibitem{10.1093/bioinformatics/bts440}
R.~Hilker, C.~Sickinger, C.~N. Pedersen, and J.~Stoye, ``{UniMoG—a unifying framework for genomic distance calculation and sorting based on DCJ},'' \emph{Bioinformatics}, vol.~28, no.~19, pp. 2509--2511, 2012.

\bibitem{GPU-quick1}
D.~Cederman and P.~Tsigas, ``A practical quicksort algorithm for graphics processors,'' in \emph{Algorithms - ESA 2008}, D.~Halperin and K.~Mehlhorn, Eds., 2008, pp. 246--258.

\bibitem{andersson1998sorting}
A.~Andersson, T.~Hagerup, S.~Nilsson, and R.~Raman, ``Sorting in linear time?'' \emph{Journal of Computer and System Sciences}, vol.~57, no.~1, pp. 74--93, 1998.

\bibitem{GPU-radix}
S.~Bandyopadhyay and S.~Sahni, ``{GRS} - {GPU} radix sort for multifield records,'' in \emph{HiPC}, 2010, pp. 1--10.

\bibitem{bit-RADIX-sort}
J.~Tang and X.~Zhou, ``Cardinality sorting and its bit-based operation-based optimization (in chinese),'' \emph{JOURNAL OF NANJINGUNIVERSITY OF TECHNOLOGY}, vol.~20, 2006.

\bibitem{zhu2021dlb}
X.~Zhu, Q.~Zhang, T.~Cheng, L.~Liu, W.~Zhou, and J.~He, ``Dlb: deep learning based load balancing,'' in \emph{CLOUD}, 2021.

\bibitem{DBLP:journals/access/XiangZCCLZ19}
W.~Xiang, H.~Zhang, R.~Cui, X.~Chu, K.~Li, and W.~Zhou, ``Pavo: {A} rnn-based learned inverted index, supervised or unsupervised?'' \emph{{IEEE} Access}, vol.~7, pp. 293--303, 2019.

\bibitem{TimKraska}
T.~Kraska, A.~Beutel, E.~H. Chi, J.~Dean, and N.~Polyzotis, ``The case for learned index structures,'' in \emph{SIGMOD}, 2018, pp. 489--504.

\bibitem{DBLP:conf/cidr/KraskaABCKLMMN19}
T.~Kraska, M.~Alizadeh, A.~Beutel, E.~H. Chi, A.~Kristo, G.~Leclerc, S.~Madden, H.~Mao, and V.~Nathan, ``Sagedb: {A} learned database system,'' in \emph{CIDR}, 2019.

\bibitem{SageDB}
J.~Ding, R.~Marcus, A.~Kipf, V.~Nathan, A.~Nrusimha, K.~Vaidya, A.~van Renen, and T.~Kraska, ``Sagedb: An instance-optimized data analytics system,'' \emph{PVLDB}, vol.~15, no.~13, 2022.

\bibitem{quick-draw}
Google, ``Google creative lab,'' Available: \url{https://github.com/googlecreativelab}, google Creative Lab [Online].

\bibitem{introcution-to-algorithm}
T.~H. Cormen, \emph{Introduction to Algorithms, 3rd Edition}.\hskip 1em plus 0.5em minus 0.4em\relax Press.

\bibitem{C++}
``C++ resources network,'' \url{http://www.cplusplus.com/}, general information about the C++ programming language, including non-technical documents and descriptions.

\bibitem{Redis}
``Redis,'' \url{https://redis.io/}, redis is an open source (BSD licensed), in-memory data structure store, used as a database, cache and message broker.

\bibitem{adam}
D.~P. Kingma and J.~Ba, ``Adam: {A} method for stochastic optimization,'' in \emph{{ICLR }}, 2015.

\bibitem{Huber1964Robust}
P.~J. Huber, ``Robust estimation of a location parameter,'' \emph{Annals of Mathematical Statistics}, vol.~35, no.~1, pp. 73--101, 1964.

\bibitem{DBLP:conf/alenex/EdelkampW19}
S.~Edelkamp and A.~Wei{\ss}, ``Worst-case efficient sorting with quickmergesort,'' in \emph{ALENEX}, 2019, pp. 1--14.

\bibitem{Python}
``Python resources network,'' \url{https://www.python.org/}, general information about the Python programming language, including non-technical documents and descriptions.

\bibitem{DBLP:conf/csr/EdelkampW14}
S.~Edelkamp and A.~Wei{\ss}, ``Quickxsort: Efficient sorting with n logn - 1.399n + o(n) comparisons on average,'' in \emph{International Computer Science Symposium in Russia}, 2014, pp. 139--152.

\bibitem{DBLP:conf/sigmod/GalakatosMBFK19}
A.~Galakatos, M.~Markovitch, C.~Binnig, R.~Fonseca, and T.~Kraska, ``Fiting-tree: {A} data-aware index structure,'' in \emph{SIGMOD}, 2019.

\bibitem{DBLP:journals/corr/abs-1901-00902}
\BIBentryALTinterwordspacing
M.~Mitzenmacher, ``A model for learned bloom filters, and optimizing by sandwiching,'' \emph{CoRR}, vol. abs/1901.00902, 2019. [Online]. Available: \url{http://arxiv.org/abs/1901.00902}
\BIBentrySTDinterwordspacing

\end{thebibliography}
%\end{small}% \newpage
%}%EAT
%\clearpage
%\input{appendix}

\end{document}